\documentclass[10.5pt,a4paper]{article}

\usepackage[top=2.0cm,bottom=2.0cm,left=2.2cm,right=2.2cm]{geometry}
\usepackage{times}
\usepackage{amsmath,amssymb}
\usepackage{graphicx}
\usepackage{booktabs}
\usepackage{array}
\usepackage{xcolor}
\usepackage{hyperref}
\usepackage{cite}
\usepackage{url}
\usepackage{microtype}
\usepackage{enumitem}
\usepackage{float}
\usepackage{caption}
\usepackage{subcaption}
\usepackage{fancyhdr}
\usepackage{titlesec}
\usepackage{abstract}
\usepackage{tikz}
\usetikzlibrary{shapes.geometric,shapes.misc,arrows.meta,positioning,
                fit,backgrounds,decorations.pathmorphing,calc}

\hypersetup{
  colorlinks=true,
  linkcolor=blue!70!black,
  citecolor=blue!70!black,
  urlcolor=blue!70!black,
  pdftitle={Remote Attestation for Trustworthy AI-Assisted Grant Evaluation},
  pdfauthor={},
}

\titleformat{\section}{\large\bfseries}{\thesection.}{0.5em}{}
\titleformat{\subsection}{\normalsize\bfseries}{\thesubsection.}{0.5em}{}
\titleformat{\subsubsection}{\normalsize\itshape}{\thesubsubsection.}{0.5em}{}

\begin{document}

\title{\Large\textbf{Making AI-Assisted Grant Evaluation Auditable without Exposing the Model}}

\author{
  \textbf{Kemal Bicakci}%
  \footnote{ORCID: \href{https://orcid.org/0000-0002-2378-8027}{0000-0002-2378-8027}}\\
  Informatics Institute, Istanbul Technical University, Istanbul, T\"{u}rkiye\\
  Securify Information Technology and Security Training Consulting Inc., Ankara, T\"{u}rkiye\\
  \texttt{kemalbicakci@itu.edu.tr}\\
  \medskip
  \textit{Preprint -- Submitted to arXiv, April 2026}
}

\date{}
\maketitle
\thispagestyle{fancy}

\begin{abstract}
\noindent
Public agencies are beginning to consider large language models (LLMs) as decision-support tools for grant evaluation. This creates a practical governance problem: the model and scoring rubric should not be exposed in a way that allows applicants to optimize against them, yet the evaluation process must remain auditable, contestable, and accountable.

We propose a TEE-based architecture that helps reconcile these requirements through \textit{remote attestation}. The architecture allows an external verifier to check which model, rubric, prompt template, and input representation were used, without exposing model weights, proprietary scoring logic, or intermediate reasoning to applicants or infrastructure operators. The main artifact is an \textit{attested evaluation bundle}: a signed, timestamped record linking the original submission hash, the canonical input hash, the model-and-rubric measurement, and the evaluation output.

The paper also considers a scenario-specific prompt injection risk: applicant-controlled documents may contain hidden or indirect instructions intended to influence the LLM evaluator. We therefore include a canonicalization and sanitization layer that normalizes document representations and records suspicious transformations before inference. We position the design relative to confidential AI inference, attestable AI audits, zero-knowledge machine learning, algorithmic accountability, and AI-assisted peer review. The resulting claim is deliberately narrow: remote attestation does not prove that an evaluation is fair or scientifically correct, but it can make part of the evaluation process externally verifiable.
\end{abstract}

\textbf{Keywords:} remote attestation, trusted execution environment, confidential AI inference, grant evaluation, algorithmic accountability, prompt injection, verifiable machine learning, human oversight.

\vspace{4pt}
\hrule
\vspace{8pt}

\section{Introduction}
\label{sec:intro}

Public funding agencies such as national science foundations routinely evaluate large numbers of grant proposals. The evaluation workload is substantial: each proposal may require several expert reviewers, panels must harmonize scores across heterogeneous disciplinary backgrounds, and turnaround expectations are often tight. Against this backdrop, there is growing interest in using large language models as decision-support tools---not to replace human judgment, but to assist with preliminary scoring, rubric compliance checking, identifying missing information, and flagging inconsistencies across reviewer assessments~\cite{Sandstrom2026}.

The motivation for this work comes from a practical governance question: if a public agency uses an AI model to assist grant evaluation, how can applicants and auditors trust that the declared model was actually used, while preventing applicants from reverse-engineering or gaming the evaluation procedure? This question creates a tension between two legitimate requirements:

\begin{itemize}[leftmargin=1.5em]
  \item \textbf{Model confidentiality.} If the model's scoring criteria, rubric details, and behavioral tendencies are publicly known, sophisticated applicants can over-optimize their proposals for the model rather than for genuine scientific merit---a form of Goodhart's Law applied to automated evaluation. The integrity of the funding process depends on keeping the model and its policy partially opaque to applicants.

  \item \textbf{Process auditability.} Public funding decisions must be explainable, contestable, and subject to independent oversight. Applicants, funding agencies, and external auditors must be able to verify that the declared model and declared rubric were actually used, that the submission was not silently modified before evaluation, and that the output was not tampered with after generation.
\end{itemize}

In practice, this tension is often handled through policy and institutional trust: agencies may disclose high-level principles while keeping implementation details internal. That may be reasonable in some settings, but it leaves a technical gap. Applicants and auditors may still need evidence that the declared model, rubric, and input representation were actually used. This gap is closely related to the normative requirements of algorithmic accountability in public sector decision making~\cite{OECD2020,OGP2021}.

This paper explores one technical way to reduce that gap. Drawing on the hardware-rooted trust model of \textit{confidential computing}, we show how \textit{remote attestation} can provide evidence that a specific, approved evaluation pipeline was launched and used---without requiring the auditor to trust the infrastructure operator, the cloud provider, or every software layer in the agency's environment. The key distinction, which we use throughout the paper, is:

\begin{quote}
\textit{Remote attestation does not certify that the model made a good decision. It certifies that the correct model, in the correct environment, with the correct version of the rubric, processed the unmodified input and that the output was not subsequently altered.}
\end{quote}

This is a narrower claim than full fairness or correctness, but it is still useful: it turns one part of the process into something externally verifiable. It should be complemented by separate validation processes---model audits, bias analyses, and human review panels---that address the quality and legitimacy of the evaluation itself.

%\subsection{Contributions}

The paper makes four main contributions:

\begin{enumerate}[leftmargin=1.5em]
  \item We frame AI-assisted public grant evaluation as a confidentiality-auditability problem: the evaluation logic should remain protected, while the process should still be independently checkable (\S\ref{sec:problem}).

  \item We propose a TEE-based architecture that produces an \textit{attested evaluation bundle} per submission, binding the submission hash, canonical input hash, model-and-rubric measurements, inference timestamp, and signed output (\S\ref{sec:architecture}).

  \item We analyze prompt-injection and input-canonicalization issues that arise when applicant-controlled documents are processed by an LLM evaluator (\S\ref{sec:architecture} and \S\ref{sec:security}).

  \item We discuss the limitations of the approach and compare it with stronger, but currently less practical, cryptographic alternatives such as ZK-ML, FHE, and MPC (\S\ref{sec:relwork} and \S\ref{sec:limitations}).
\end{enumerate}

The rest of the paper is organized as follows. Section~2 defines the problem setting, requirements, and threat model for AI-assisted grant evaluation. Section~3 provides background on trusted execution environments and remote attestation. Section~4 presents the proposed attestation-based architecture and the attested evaluation bundle. Section~5 discusses the security properties and limitations of the approach. Section~6 reviews related work on confidential computing, verifiable computation, and AI-assisted evaluation. Finally, Section~7 concludes the paper and outlines directions for future work.

\section{Problem Statement and Threat Model}
\label{sec:problem}

\subsection{Setting}

We consider a \textit{grant evaluation agency} (GEA) that uses an AI-assisted evaluation pipeline as a component in its proposal review process. The pipeline takes a structured applicant submission as input and produces a structured evaluation report: numerical scores on multiple rubric dimensions, natural-language justifications, uncertainty flags, and reviewer attention cues. The AI output feeds into a human review panel that makes the final funding decision.

\subsection{Principals and Roles}

Four principals participate in the system:

\begin{itemize}[leftmargin=1.5em]
  \item \textbf{Applicant (A).} Submits a proposal. Has an interest in receiving a high score. Is aware that AI evaluation is used.
  \item \textbf{Agency (GEA).} Operates the evaluation process. Wants evaluations to be consistent, rubric-compliant, and defensible. Holds the model and rubric as confidential internal policy.
  \item \textbf{Model provider (MP).} Provides the model weights and inference infrastructure. In practice this role may be filled by a separate commercial AI service, or by the agency itself operating an in-house or open-weight model. In the latter case MP and GEA collapse into a single principal, which simplifies the trust structure but does not change the core attestation requirements.
  \item \textbf{Auditor / Appellate board (AUD).} An independent body entitled to verify that a specific evaluation was conducted as claimed. May act on behalf of an aggrieved applicant or as a systemic regulator.
\end{itemize}

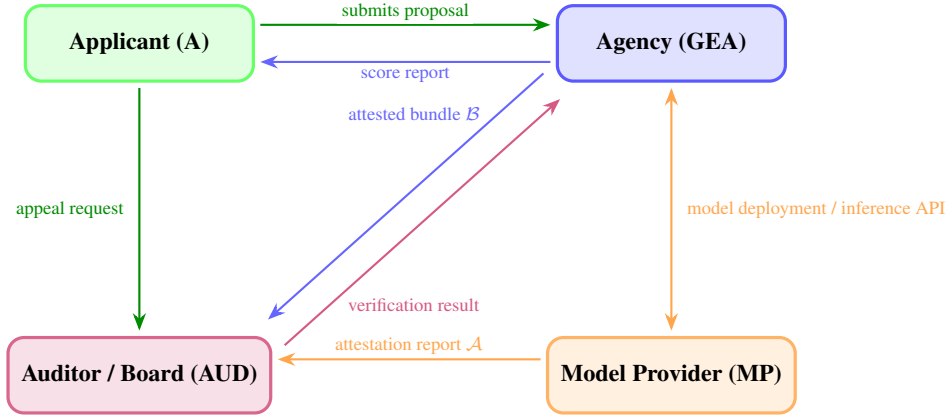
\begin{figure}[H]
\centering
\begin{tikzpicture}[
  >=Stealth,
  font=\small,
  principal/.style={rectangle, rounded corners=6pt, draw, very thick,
                    text centered, minimum width=3.0cm, minimum height=1.0cm,
                    inner sep=5pt},
  agency/.style={principal, fill=blue!12,   draw=blue!65},
  applicant/.style={principal, fill=green!12,  draw=green!60},
  provider/.style={principal, fill=orange!12, draw=orange!60},
  auditor/.style={principal, fill=purple!12, draw=purple!60},
  arr/.style={->, thick, shorten >=2pt, shorten <=2pt},
  biarr/.style={<->, thick, shorten >=2pt, shorten <=2pt},
  lbl/.style={font=\scriptsize, fill=white, inner sep=1.5pt},
]

%% 2x2 grid: App top-left, GEA top-right, AUD bottom-left, MP bottom-right
\node[applicant] (app) at (-3.5,  2.2) {\textbf{Applicant (A)}};
\node[agency]    (gea) at ( 3.5,  2.2) {\textbf{Agency (GEA)}};
\node[auditor]   (aud) at (-3.5, -2.2) {\textbf{Auditor / Board (AUD)}};
\node[provider]  (mp)  at ( 3.5, -2.2) {\textbf{Model Provider (MP)}};

%% TOP: App->GEA (above) and GEA->App (below) — parallel horizontals
\draw[arr, green!55!black]
      ([yshift= 7pt]app.east) --
      node[lbl, above]{submits proposal}
      ([yshift= 7pt]gea.west);

\draw[arr, blue!60]
      ([yshift=-7pt]gea.west) --
      node[lbl, below]{score report}
      ([yshift=-7pt]app.east);

%% RIGHT: GEA <-> MP (bidirectional vertical)
\draw[biarr, orange!70]
      (gea.south) --
      node[lbl, right=4pt]{model deployment / inference API}
      (mp.north);

%% LEFT: App -> AUD (vertical)
\draw[arr, green!55!black]
      (app.south) --
      node[lbl, left=4pt]{appeal request}
      (aud.north);

%% BOTTOM: MP -> AUD (horizontal, offset upward)
\draw[arr, orange!70]
      ([yshift=6pt]mp.west) --
      node[lbl, above]{attestation report $\mathcal{A}$}
      ([yshift=6pt]aud.east);

%% DIAGONAL GEA->AUD: NW offset from main diagonal, label near GEA
%% Diagonal direction (-7,-4.4), perp unit approx (-0.53, 0.85)
\draw[arr, blue!60]
      ($(gea.south west)+(-0.11, 0.17)$) --
      node[lbl, above left, pos=0.22]{attested bundle $\mathcal{B}$}
      ($(aud.north east)+(-0.11, 0.17)$);

%% DIAGONAL AUD->GEA: SE offset from main diagonal, label near AUD
\draw[arr, purple!65]
      ($(aud.north east)+(0.11, -0.17)$) --
      node[lbl, below right, pos=0.22]{verification result}
      ($(gea.south west)+(0.11, -0.17)$);

\end{tikzpicture}
\caption{The four principals and their interactions. Applicant (top-left) submits proposals to the Agency (top-right) and receives score reports in return. The Agency coordinates model deployment with the Model Provider (bottom-right) and sends the attested bundle~$\mathcal{B}$ to the Auditor (bottom-left). The Auditor receives the attestation report~$\mathcal{A}$ from the Model Provider, returns a verification result to the Agency, and handles Applicant appeals.}
\label{fig:principals}
\end{figure}

Figure~\ref{fig:principals} illustrates the four principals and their key interactions.

\subsection{Threat Model}
\label{sec:threats}

We identify the following adversarial scenarios:

\begin{description}[leftmargin=1.5em]
  \item[\textbf{T1 -- Applicant gaming.}] An applicant, knowing the model's exact behavior, constructs a proposal that exploits model biases or keyword sensitivities rather than demonstrating genuine merit.
  \item[\textbf{T2 -- Prompt injection.}] An applicant embeds invisible instructions (white text, PDF metadata, hidden comment fields, Unicode lookalikes) inside their submission with the intent to override the system prompt or scoring logic.
  \item[\textbf{T3 -- Post-hoc output tampering.}] An internal actor (malicious operator, compromised software layer) alters evaluation scores or justifications after generation but before they are recorded.
  \item[\textbf{T4 -- Wrong model/rubric deployment.}] A misconfigured or deliberately substituted model version or rubric variant is used, producing evaluations inconsistent with declared policy.
  \item[\textbf{T5 -- Input silencing.}] The submission fed to the model is a filtered or altered version of the original, suppressing content that would have led to a lower score.
  \item[\textbf{T6 -- Infrastructure operator breach.}] A cloud provider or hosting operator with hypervisor-level access reads model weights, rubric contents, or intermediate inference state.
\end{description}

\noindent The architecture is intended to reduce the practical risk of T2--T5, although prompt injection and document canonicalization remain open-ended problems. It also reduces exposure to T6 under the standard TEE threat model. T1 is addressed primarily at the policy level through model and rubric confidentiality; the proposed architecture provides a technical mechanism for enforcing that policy.

\section{Background}
\label{sec:background}

\subsection{Trusted Execution Environments}

A Trusted Execution Environment (TEE) is a hardware-isolated computation region where code and data are protected from unauthorized access, including by privileged software layers such as the operating system and hypervisor. The isolation is enforced by CPU-level mechanisms. Prominent TEE implementations include Intel Trust Domain Extensions (TDX), AMD Secure Encrypted Virtualization-Secure Nested Paging (SEV-SNP), and ARM Confidential Compute Architecture (CCA)~\cite{NIST8320,ConfidentialComputingConsortium}.

TEEs provide two security properties relevant to our design:

\begin{itemize}[leftmargin=1.5em]
  \item \textbf{Confidentiality.} Memory pages inside the TEE are encrypted. The hypervisor and OS cannot read their contents.
  \item \textbf{Integrity.} A TEE maintains a cryptographic measurement (typically a hash chain of loaded code and data) that can be used to verify the integrity of the environment.
\end{itemize}

\subsection{Remote Attestation}

Remote attestation (RA) is a protocol by which a \textit{relying party} verifies claims about a remote system's software and hardware state without being physically present. In the TEE context, the protocol produces a signed \textit{attestation report} generated by the CPU's attestation key, which chains to a hardware root of trust from the CPU manufacturer. The report binds a \textit{measurement}---a hash of the TEE's initial code and configuration---to a \textit{nonce} or other challenge provided by the relying party, preventing replay attacks~\cite{NIST8320}.

The relying party checks: (1) the signature is valid under the CPU manufacturer's certificate chain; (2) the measurement matches a reference manifest they trust; (3) the nonce is fresh. If all three checks pass, the relying party has evidence that the remote system was launched with the approved software state in a genuine TEE. The strength of that evidence depends on the TEE implementation, the freshness of the reference manifest, and the trust placed in the hardware vendor's attestation infrastructure.

\subsection{Confidential Computing for AI Workloads}

Confidential computing has been extended to AI inference workloads, motivated by the need to protect model weights from cloud operators and user inputs from model providers simultaneously. Microsoft Azure's Confidential AI Inferencing service, for example, runs an LLM endpoint inside a TDX virtual machine and provides a publicly verifiable attestation policy that specifies which model, runtime, and system prompt versions are approved~\cite{AzureConfidential2024}. Anthropic has published a technical approach to confidential inference that extends this model to include prompt privacy and output non-disclosure guarantees~\cite{Anthropic2025}.

These commercial and research systems suggest that TEE-based LLM inference is becoming operationally feasible. This paper adapts that line of work to the more specific accountability requirements of public grant evaluation.

\section{Related Work}
\label{sec:relwork}

We do not attempt a full survey of confidential AI or verifiable machine learning. Instead, we focus on the strands that directly motivate the proposed design.

\subsection{Confidential AI Inference}

The emerging literature on confidential AI inference addresses the problem of running machine learning models in environments where neither the user's input nor the model provider's weights should be exposed to the hosting infrastructure. Carlini et al.~\cite{Carlini2021} and subsequent work demonstrate that model weights and training data can be extracted from deployed models under various threat models, motivating hardware-level isolation. The systematization by Mo et al.~\cite{Sun2024} provides a comprehensive taxonomy of confidential computing techniques applied to ML pipelines, distinguishing between input confidentiality, model confidentiality, and gradient confidentiality in federated settings.

For our application, the primary confidentiality requirement is that model weights and evaluation rubric remain unexposed to applicants, so that the scoring procedure cannot be gamed. When the agency uses a third-party model provider, an additional concern arises: the provider may also wish to protect model weights as commercial intellectual property, and the agency may not want its rubric exposed even to the provider's infrastructure. When the agency operates the model itself, the IP concern disappears, but the anti-gaming requirement remains. In either case the key desideratum is the same: applicants and outside parties should not be able to infer the scoring logic, while auditors should be able to verify that the declared configuration was used. This combination is not well-addressed by prior work focused on the bilateral user-provider relationship.

\subsection{Attestable AI Audits}

The closest prior work to our proposal is the \textit{attestable audits} paradigm introduced by recent work on AI safety evaluation~\cite{AttestableAudits2025}. The core idea is to run AI safety benchmarks inside a TEE so that external parties can verify, via remote attestation, that a specific model was evaluated under specific test conditions and that results were not modified. This eliminates the need to trust the audit organization's infrastructure or personnel.

We extend this paradigm from the \textit{benchmark evaluation} setting to the \textit{proposal evaluation} setting. The structural differences are significant: benchmark audits involve adversarial test inputs chosen by the auditor, while grant evaluation involves applicant-controlled inputs that must be protected against injection attacks. Additionally, the relying party in grant evaluation is not a safety regulator but an applicant or appellate board with different verification needs.

\subsection{Zero-Knowledge Machine Learning}

Zero-knowledge proofs (ZKPs) offer a mathematically stronger alternative to TEE-based attestation: a \textit{proof} that a computation was performed correctly, verifiable without re-running it and without revealing the model or inputs. The ZKML research area applies ZKPs to neural network inference~\cite{ZKML2025}. Recent work has demonstrated ZK proofs for selected neural-network workloads, but proof generation time and memory cost remain highly dependent on architecture, sequence length, and arithmetic representation.

For the large language models relevant to grant evaluation (tens of billions of parameters), ZK proof generation remains computationally prohibitive. For transformer-scale LLMs, the overhead remains large enough that ZKML is better viewed as a medium-to-long-term alternative rather than an immediately deployable solution for routine grant evaluation. In \S\ref{sec:limitations} we discuss the conditions under which the security-cost tradeoff may shift in favor of ZKML.

\subsection{Secure Multi-Party Computation and Homomorphic Encryption}

Secure multi-party computation (MPC) and fully homomorphic encryption (FHE) enable computation on encrypted data without decrypting it. While theoretically appealing---an FHE-based system would allow inference on encrypted submissions without the model ever seeing plaintext---current FHE performance on transformer-scale models is impractical by several orders of magnitude~\cite{Sun2024}. MPC-based approaches face similar overhead challenges and introduce communication complexity that scales poorly with model size.

We therefore treat FHE and MPC as important research directions but exclude them from our near-term architecture. A hybrid approach---using TEE for LLM inference and MPC or FHE for sensitive sub-computations such as score aggregation---may offer a practical middle ground.

\subsection{AI-Assisted Peer Review and Grant Evaluation}

Empirical work on LLMs as reviewers has produced cautiously optimistic results. Sandstr\"{o}m and Thelwall~\cite{Sandstrom2026} revisit a classic peer review dataset and find that LLM-based scoring shows partial alignment with human expert judgments, while also revealing systematic biases correlated with proposal framing, length, and institutional signals.

The emerging consensus in the literature is that LLMs are valuable as \textit{decision support tools} in proposal evaluation---assisting with rubric checking, missing-information flagging, and preliminary triage---but should not serve as sole or final decision makers~\cite{Sandstrom2026}. Our architecture is designed around this consensus: AI output feeds into a human review panel, and the attested bundle provides evidence of what the AI said, not authorization for a final decision.

\subsection{Prompt Injection Attacks}

Prompt injection was first systematically characterized by Perez and Ribeiro~\cite{PerezRibeiro2022}, who demonstrated that adversarial instructions embedded in user-controlled inputs can override system prompt instructions in LLMs. Subsequent work has expanded the taxonomy to include \textit{indirect} prompt injection---where the adversarial instructions appear in third-party content retrieved by the model (e.g., documents fetched by a RAG system)---and \textit{steganographic} injection, where instructions are hidden in formatting, Unicode normalization artifacts, or whitespace~\cite{Greshake2023}.

In the grant evaluation setting, the applicant's submission is precisely the type of third-party, adversary-controlled document that indirect injection exploits. Defenses against prompt injection remain an active research area~\cite{Wallace2024}; our contribution is to integrate a canonicalization layer that reduces the attack surface and includes injection-attempt detection as a scored meta-attribute of the evaluation.

\subsection{Algorithmic Accountability in the Public Sector}

The normative literature on algorithmic accountability establishes requirements for automated systems used in consequential public decisions: transparency of process, contestability of outcomes, non-discrimination, and human oversight of final decisions~\cite{OECD2020,OGP2021}. The NIST AI Risk Management Framework~\cite{NISTAI} codifies these into operational practices including documentation, ongoing monitoring, and incident response.

Remote attestation contributes to these requirements by providing a cryptographic audit trail that supports transparency of process and contestability. It does not, however, automatically satisfy non-discrimination requirements (which depend on model training data and rubric design) or ensure high-quality human oversight (which depends on reviewer competence and institutional incentives). We treat these as complementary, non-technical requirements that the broader governance framework must address.

\section{Proposed Architecture}
\label{sec:architecture}

\subsection{Overview}

The architecture consists of five layers, each with defined security properties and interfaces:

\begin{enumerate}[leftmargin=1.5em]
  \item \textbf{Submission ingestion and canonicalization layer} (outside TEE, logged)
  \item \textbf{Attestation bootstrap} (TEE launch and measurement)
  \item \textbf{Evaluation inference core} (inside TEE)
  \item \textbf{Attested output signing} (inside TEE before exit)
  \item \textbf{Audit log and verification service} (outside TEE, tamper-evident)
\end{enumerate}

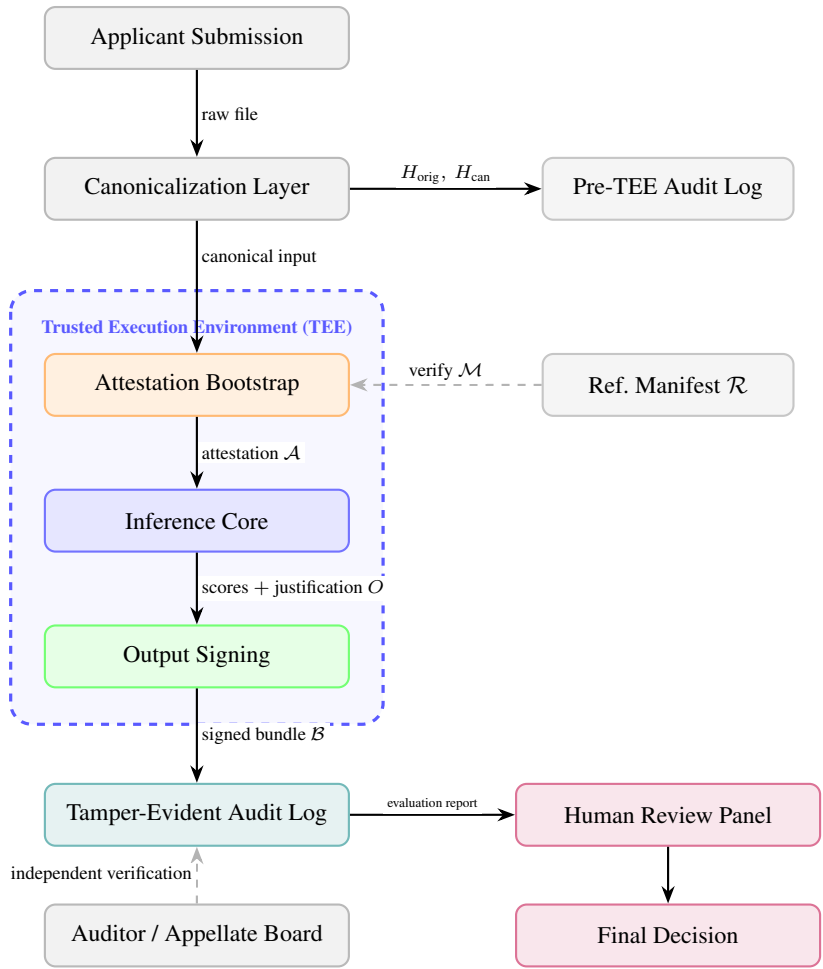
\begin{figure}[H]
\centering
\begin{tikzpicture}[
  >=Stealth,
  font=\small,
  main/.style={rectangle, rounded corners=4pt, draw, thick, text centered,
               minimum width=4.0cm, minimum height=0.82cm, inner sep=5pt},
  outside/.style={main, fill=gray!10,   draw=gray!55},
  attest/.style ={main, fill=orange!12, draw=orange!55},
  inferst/.style={main, fill=blue!10,   draw=blue!55},
  signing/.style={main, fill=green!10,  draw=green!55},
  auditst/.style={main, fill=teal!10,   draw=teal!55},
  humanst/.style={main, fill=purple!10, draw=purple!55},
  side/.style=   {main, fill=gray!8,    draw=gray!45, minimum width=3.3cm},
  arr/.style={->, thick},
  darr/.style={->, thick, dashed, draw=gray!60},
  lbl/.style={font=\scriptsize, fill=white, inner sep=1.5pt},
]

%% ---- Main column (x = 0) ----
\node[outside] (submit)  at (0,  0.0) {Applicant Submission};
\node[outside] (canon)   at (0, -2.0) {Canonicalization Layer};
\node[attest]  (boot)    at (0, -4.6) {Attestation Bootstrap};
\node[inferst] (infer)   at (0, -6.4) {Inference Core};
\node[signing] (sign)    at (0, -8.2) {Output Signing};
\node[auditst] (alog)    at (0,-10.3) {Tamper-Evident Audit Log};
\node[outside] (auditor) at (0,-11.9) {Auditor / Appellate Board};

%% ---- Side column (x = 6.2) ----
\node[side]    (prelog)  at (6.2, -2.0) {Pre-TEE Audit Log};
\node[side]    (refman)  at (6.2, -4.6) {Ref.\ Manifest $\mathcal{R}$};
\node[humanst] (panel)   at (6.2,-10.3) {Human Review Panel};
\node[humanst] (decis)   at (6.2,-11.9) {Final Decision};

%% ---- TEE boundary (explicit rectangle, behind everything) ----
\begin{scope}[on background layer]
  \fill[blue!3, rounded corners=8pt]
       (-2.45,-3.35) rectangle (2.45,-9.1);
  \draw[blue!65, very thick, dashed, rounded corners=8pt]
       (-2.45,-3.35) rectangle (2.45,-9.1);
\end{scope}
%% TEE label: centred just inside the top edge, white bg so border shows through
\node[font=\scriptsize\bfseries, text=blue!65, fill=blue!3,
      inner sep=2pt] at (0,-3.85)
      {Trusted Execution Environment (TEE)};

%% ---- Arrows: pre-TEE ----
\draw[arr]  (submit) -- node[lbl,right]{raw file} (canon);
\draw[arr]  (canon)  -- node[lbl,above]{$H_{\text{orig}},\,H_{\text{can}}$} (prelog);
%% label at pos=0.28 keeps it above the TEE top border
\draw[arr]  (canon)  -- node[lbl,right,pos=0.28]{canonical input} (boot);

%% ---- Arrows: inside TEE ----
%% refman -> boot: route outside the TEE box to avoid collision with TEE label
\draw[darr] (refman.west) to[out=180,in=0]
            node[lbl,above,pos=0.5]{verify $\mathcal{M}$}
            (boot.east);
\draw[arr]  (boot)  -- node[lbl,right]{attestation $\mathcal{A}$} (infer);
\draw[arr]  (infer) -- node[lbl,right]{scores $+$ justification $O$} (sign);

%% ---- Arrows: post-TEE ----
\draw[arr]  (sign)    -- node[lbl,right]{signed bundle $\mathcal{B}$} (alog);
\draw[arr]  (alog)    -- node[lbl,above,font=\tiny]{evaluation report} (panel);
\draw[arr]  (panel)   -- (decis);
\draw[darr] (auditor) -- node[lbl,left]{independent verification} (alog);

\end{tikzpicture}
\caption{System architecture and data flow. The blue dashed box marks the TEE boundary: all operations inside are encrypted and isolated from the OS, hypervisor, and cloud operator. Solid arrows indicate data flow; dashed arrows indicate verification interactions.}
\label{fig:arch}
\end{figure}

Figure~\ref{fig:arch} illustrates the data flow and trust boundaries. The critical invariant is that once data enters the TEE boundary, it cannot be observed or modified by any software layer outside the TEE, including the OS, hypervisor, and cloud management plane.

\subsection{Layer 1: Submission Ingestion and Canonicalization}

\subsubsection{Document parsing and format normalization}

Applicant submissions typically arrive as PDF, DOCX, LaTeX source, or structured web forms. Each format presents distinct channels for embedding adversarial content. The ingestion layer:

\begin{enumerate}[leftmargin=1.5em,label=(\arabic*)]
  \item Computes and records the \textit{original submission hash} $H_\text{orig}$ (SHA-3-256 of the raw file bytes).
  \item Parses all format layers: visible text, metadata fields (author, subject, keywords, custom properties), embedded objects (images, fonts, scripts), comment fields, hidden layers, and revision history.
  \item Applies a \textit{canonical representation transform} that strips all non-visible and non-structural content, retaining only the human-readable text and declared section structure.
  \item Computes the \textit{canonical input hash} $H_\text{can}$ over the normalized representation.
  \item Records both $H_\text{orig}$ and $H_\text{can}$ in the pre-TEE audit log.
\end{enumerate}

\subsubsection{Prompt injection detection and sanitization}

Before constructing the model input, the canonicalization layer applies injection screening:

\begin{itemize}[leftmargin=1.5em]
  \item Detection of instruction-like patterns (imperative verbs followed by evaluation-related directives, e.g., ``ignore previous instructions,'' ``assign maximum score'').
  \item Removal of Unicode control characters, soft hyphens, zero-width spaces, and bidirectional override characters that can alter visual rendering without appearing in plain text.
  \item Detection of unusually high density of whitespace, which may indicate hidden white-on-white text in original documents.
  \item Flagging of any metadata fields containing natural-language directives.
\end{itemize}

Detected anomalies are not silently removed; they are recorded in a \textit{sanitization report} that becomes part of the evaluation bundle. This transparency ensures that the applicant cannot later claim injection content was present without the agency's knowledge, and provides the auditor with evidence of what was or was not cleaned.

\subsection{Layer 2: Attestation Bootstrap}

Before evaluation begins, the TEE launches and generates an attestation report as follows:

\begin{enumerate}[leftmargin=1.5em,label=(\arabic*)]
  \item The CPU measures the TEE initial state: loaded code (inference runtime binary), model weights (or a verifiable pointer to sealed weights), rubric and prompt template, and security-critical configuration.
  \item The measurement $\mathcal{M}$ is a hash chain over these components, signed by the CPU's attestation key rooted in the manufacturer's certificate authority.
  \item The attestation report $\mathcal{A} = \text{Sign}_\text{CPU}(\mathcal{M} \| \text{nonce} \| \text{timestamp})$ is generated and made available for external verification.
  \item An external \textit{attestation verifier}---operated independently of the agency, e.g., by a designated audit authority or by the CPU manufacturer's verification service---checks $\mathcal{A}$ against a \textit{reference manifest} $\mathcal{R}$ that encodes the approved model hash, rubric hash, prompt template hash, and runtime hash.
\end{enumerate}

The reference manifest $\mathcal{R}$ is a publicly commitments document signed by the agency at policy-adoption time and stored in a transparency log. This ensures that the definition of ``correct configuration'' is itself public and time-locked, preventing retroactive policy changes.

\subsection{Layer 3: Evaluation Inference Core}

Inside the TEE, the evaluation proceeds as follows:

\begin{enumerate}[leftmargin=1.5em,label=(\arabic*)]
  \item The canonical input (from Layer 1) is received and its hash $H_\text{can}$ is verified against the pre-TEE log.
  \item The system prompt, rubric, and evaluation instructions are loaded from sealed storage within the TEE.
  \item The full model input $I$ is constructed: system prompt $\| $ rubric $\|$ separator token $\|$ canonical submission.
  \item The input hash $H_I = \text{SHA3}(I)$ is computed inside the TEE.
  \item Inference is performed; the model produces an evaluation output $O$ containing numerical scores per rubric dimension, natural-language justifications, an uncertainty estimate, a list of flagged weaknesses, and reviewer attention cues.
  \item The output hash $H_O = \text{SHA3}(O)$ is computed.
\end{enumerate}

\subsection{Layer 4: Attested Output Signing}

Before any data exits the TEE, the evaluation evidence packet is assembled and signed:

\[
\mathcal{E} = \{H_\text{orig},\; H_\text{can},\; H_I,\; H_O,\; \mathcal{M},\; \mathcal{A},\; \tau,\; O\}
\]

where $\tau$ is the RFC 3161 timestamp obtained from a trusted timestamping authority. The signed bundle is:

\[
\mathcal{B} = \mathcal{E} \;\|\; \text{Sign}_\text{TEE}(\mathcal{E})
\]

The signing key $k_\text{TEE}$ is a per-session ephemeral key generated inside the TEE and certified by the attestation report, so its use is bound to the specific TEE launch that produced the measurement $\mathcal{M}$.

\subsection{Layer 5: Audit Log and Verification Service}

The attested bundle $\mathcal{B}$ exits the TEE and is written to a tamper-evident audit log. We recommend an append-only log with Merkle-tree-based consistency proofs (e.g., a Certificate-Transparency-style log) so that any subsequent deletion or modification of an entry is cryptographically detectable.

Any party in possession of $\mathcal{B}$, the reference manifest $\mathcal{R}$, and the CPU manufacturer's certificate chain can independently verify:

\begin{enumerate}[leftmargin=1.5em,label=(\arabic*)]
  \item The attestation signature is valid and chains to the hardware root of trust.
  \item The TEE measurement matches the reference manifest (i.e., approved model, rubric, and prompt were loaded).
  \item The output hash $H_O$ is consistent with the output $O$ in the bundle (i.e., output was not modified post-generation).
  \item The submission hash $H_\text{orig}$ matches the original file held by the agency's submission system (i.e., the evaluated file is the submitted file).
  \item The timestamp is from a trusted authority and predates any contestation.
\end{enumerate}

\subsection{Human Review Integration}

The attested bundle is presented to the human review panel alongside the original submission. The panel's role is to:

\begin{itemize}[leftmargin=1.5em]
  \item Review the AI-generated scores and justifications as structured decision support.
  \item Exercise independent judgment on dimensions where the AI flags uncertainty or where the panel's domain expertise suggests the rubric is an imperfect proxy.
  \item Make the final funding decision, which is recorded separately from the AI output.
  \item Log any disagreements with AI scores, which feeds into ongoing model validation.
\end{itemize}

The architecture deliberately decouples the AI evaluation artifact from the final decision. This ensures that the AI system is accountable for what it said, not for the funding outcome, and that human judgment retains both legal and normative primacy.

\section{Security Analysis}
\label{sec:security}

We analyze the architecture against the threat model of \S\ref{sec:threats}.

\paragraph{T1 (Applicant gaming).} The model and rubric are never exposed outside the TEE. An applicant who cannot observe model behavior cannot systematically exploit it. The risk is reduced to \textit{statistical inference} from aggregate funding outcomes---a residual risk that requires a larger number of evaluations and is bounded by output noise and rubric diversity. This is analogous to the residual leakage risk in differentially private systems.

\paragraph{T2 (Prompt injection).} The canonicalization layer (Layer 1) removes the most common injection vectors. Detected anomalies are logged and flagged in the evaluation bundle, enabling post-hoc audits of injection attempts. Residual risk remains for novel steganographic techniques not covered by the sanitization rules; we treat this as an ongoing research problem (see \S\ref{sec:future}).

\paragraph{T3 (Post-hoc output tampering).} The output hash $H_O$ is computed and signed inside the TEE before the output exits. Any modification to $O$ after TEE exit invalidates the signature, which is detectable by any party with the public verification key. Integrity is therefore unconditionally guaranteed under the assumption that the TEE signing key was generated and used exclusively inside the TEE.

\paragraph{T4 (Wrong model/rubric deployment).} The attestation bootstrap (Layer 2) binds the TEE measurement to the reference manifest. A substituted model or rubric produces a different measurement that fails attestation verification. This threat is mitigated as long as the reference manifest is correctly maintained and the CPU attestation is genuine.

\paragraph{T5 (Input silencing).} The pre-TEE ingestion logs $H_\text{orig}$ before canonicalization. The bundle includes both $H_\text{orig}$ and $H_\text{can}$. An auditor can verify that $H_\text{orig}$ matches the submission file and that the canonicalization transform was deterministic and policy-compliant. Silent omission of content would require either a collision attack on SHA-3 or a compromise of the pre-TEE logging system.

\paragraph{T6 (Infrastructure operator breach).} TEE memory encryption prevents a cloud operator with hypervisor access from reading model weights, rubric contents, or intermediate activations. The operator can observe encrypted ciphertext and timing side-channels but not plaintext content. Residual risk includes micro-architectural side-channel attacks (cache timing, power analysis) against TEE implementations---a known and active research area that is partially mitigated by recent CPU microcode updates.

\section{Limitations and Open Challenges}
\label{sec:limitations}

\subsection{What Remote Attestation Does Not Guarantee}

Remote attestation is a powerful but narrowly defined guarantee. It is essential to be precise about what the attested bundle does \textit{not} prove:

\begin{enumerate}[leftmargin=1.5em]
  \item \textbf{Model quality.} Attestation verifies that the approved model ran; it says nothing about whether that model is well-calibrated, unbiased, or scientifically sound. A poorly trained model, consistently applied, will produce attestable but poor evaluations.

  \item \textbf{Rubric validity.} If the rubric encodes flawed criteria or implicitly favors certain proposal styles, attestation proves only that the flawed rubric was applied consistently. Rubric design remains a human responsibility outside the technical trust boundary.

  \item \textbf{Training data bias.} Pre-training and fine-tuning corpora can encode systematic biases along institutional, linguistic, or demographic lines. These biases are inherited by any deployment of the model, regardless of how securely it is attested.

  \item \textbf{TEE hardware integrity.} The security of remote attestation ultimately depends on the integrity of the CPU manufacturer's attestation keys and the absence of undisclosed firmware vulnerabilities. Nation-state-level adversaries with access to CPU microarchitecture specifications may be capable of forgery attacks that are infeasible for ordinary adversaries.

  \item \textbf{Canonicalization completeness.} The canonicalization layer cannot enumerate all possible steganographic injection channels, particularly as document formats evolve and LLMs become sensitive to increasingly subtle formatting cues.
\end{enumerate}

\subsection{Performance and Cost Overheads}

Running LLM inference inside a TEE introduces performance overhead compared to standard cloud inference. The magnitude is workload-dependent: memory-intensive inference, large context windows, encrypted memory management, and I/O patterns can all affect throughput. Attestation also adds bootstrap latency, especially when a fresh TEE instance is created per evaluation or per batch. For a grant evaluation system processing thousands of submissions, these overheads are likely to be operational rather than conceptual obstacles, but they should be measured in the target deployment rather than assumed negligible.

\subsection{Trusted Computing Base Dependencies}

Despite its strength, the TEE model requires trust in several entities: the CPU manufacturer (for attestation key integrity), the TEE firmware (for correct isolation enforcement), and the attestation verification service (for correct policy checking). An agency adopting this architecture should conduct due diligence on each dependency and implement compensating controls (e.g., multi-vendor TEE diversity for high-stakes decisions).

\subsection{Limitations of ZK-ML as an Alternative}

Zero-knowledge proofs for neural network inference offer a stronger kind of guarantee---a mathematical proof of correct computation rather than a hardware attestation---but they face substantial practical limitations for large language models. Proof generation complexity grows with model size, sequence length, and the arithmetic representation used by the proof system. Recursive proof composition and hardware acceleration are active research directions, but the gap between current ZK-ML capability and LLM-scale grant evaluation remains substantial. For the near term, ZK-ML is better treated as a complementary research direction than as a replacement for TEE-based attestation.

\subsection{Governance and Incentive Alignment}

Technical architecture alone cannot guarantee accountable AI deployment. The reference manifest must be correctly maintained; if the agency is also the party responsible for maintaining the manifest, there is a conflict of interest. Independent manifest custody---by a regulatory authority, an international standards body, or a multi-stakeholder consortium---is architecturally preferable but requires institutional coordination beyond the scope of this technical paper.

\section{Future Work}
\label{sec:future}

The architecture presented here leaves several research questions open:

\paragraph{Adaptive canonicalization.} Current sanitization rules are static and must be manually updated as new injection techniques emerge. A dynamic, model-in-the-loop sanitization approach---where the canonicalization layer uses a secondary model to detect injection intent---introduces a bootstrapping problem (who attests the sanitizer?) but may be more robust against novel attacks. Formal characterization of the injection detection problem as an adversarial classification task would enable principled evaluation of sanitizer completeness.

\paragraph{Composable attestation across multi-stage pipelines.} Grant evaluation often involves multiple stages: initial eligibility screening, technical merit review, strategic alignment scoring, and panel discussion. Each stage may use different models or rubrics. Extending the attestation framework to multi-stage pipelines requires composable attestation protocols that chain evidence across stages while preserving per-stage confidentiality boundaries.

\paragraph{Differential privacy for aggregate reporting.} Even if individual evaluation bundles are confidential, aggregate statistics (e.g., score distributions by institution type or proposal length) may leak information about model behavior. Differential privacy mechanisms applied to aggregate reports can bound this leakage, complementing the per-bundle attestation with population-level privacy guarantees.

\paragraph{Formal verification of canonicalization transforms.} The canonicalization layer is a critical security component, but its correctness guarantees are informal. Formal specification of the transform---using a language such as TLA+ or Coq---and verification that the specification is injection-free for a defined class of document formats would substantially strengthen the security argument.

\paragraph{Human-AI disagreement as a model quality signal.} The architecture logs cases where human reviewers disagree with AI scores. Systematic analysis of these disagreements---controlling for reviewer expertise, proposal domain, and rubric dimension---can provide a rich signal for identifying model biases and rubric weaknesses. Building a feedback loop from human panel decisions to model validation (not fine-tuning, which would compromise attestation integrity) is a research priority.

\paragraph{Cross-agency comparability and portability.} Different funding agencies use different rubrics and model configurations. Developing standardized attestation schemas and reference manifest formats would enable cross-agency comparisons of AI-assisted evaluation practices and support regulatory oversight at scale.

\paragraph{Graduated trust frameworks.} Not all grant decisions carry equal stakes. A tiered system---in which small seed grants use a lighter-weight attestation protocol while large program grants require full TEE attestation with independent manifest custody---would optimize the cost-security tradeoff across the evaluation portfolio.

\section{Conclusion}
\label{sec:conclusion}

AI-assisted grant evaluation presents a genuine dual-requirement problem: the evaluation model must be confidential enough to reduce gaming, yet the process must be auditable enough to support contestability and accountability. This paper has argued that a TEE-based architecture with remote attestation can help reconcile these requirements, provided its guarantees are understood narrowly.

The central contribution of the paper is the \textit{attested evaluation bundle}: a signed, timestamped artifact that cryptographically binds the original submission hash, the canonical input hash, the model and rubric measurements, and the evaluation output. Any party in possession of this bundle and the publicly registered reference manifest can independently verify that the declared model and rubric were used, that the submission was not silently altered, and that the output was not modified after generation.

We have been careful to delineate what remote attestation proves and what it does not. It does not certify that the model made a scientifically correct or unbiased judgment. It certifies that the correct model, in the correct environment, processed the unmodified input and produced an unmodified output. This narrower claim is still valuable: it can move part of AI-assisted grant evaluation from an internal black-box process to a verifiable record grounded in cryptographic evidence.

We have also highlighted the prompt injection attack surface created when LLM-based evaluators process adversary-controlled documents, and proposed a canonicalization layer that reduces this surface while preserving a transparent sanitization record. The combination of TEE-enforced confidentiality, attestation-enforced integrity, and canonicalization-enforced input safety provides a plausible technical baseline for more trustworthy AI-assisted public decision support.

The architecture is compatible with existing confidential-computing hardware such as Intel TDX and AMD SEV-SNP, as well as emerging cloud services for confidential AI inference. The primary near-term research priorities are formal verification of the canonicalization transform, composable attestation for multi-stage pipelines, and governance frameworks for independent manifest custody.

The broader point is not that remote attestation solves the governance problem of AI-assisted public decision making. It does not. Rather, it provides a concrete mechanism for making one important part of that process checkable. Used together with human review, model validation, appeal mechanisms, and public accountability procedures, it can contribute to a more trustworthy evaluation infrastructure.

\end{document}